\documentclass[prc,showpacs,showkeys,superscriptaddress,nofootinbib,twocolumn,floatfix]{revtex4}
\usepackage{amsfonts}
\usepackage{graphicx,color,amsmath,amssymb}

\def\blfootnote{\xdef\@thefnmark{}\@footnotetext}
\newcommand{\ie}{\textit{i.e.}}

\newcommand{\gsim}{\gtrsim}

\begin{document}

\title{Ideal Hydrodynamics for Bulk and Multistrange Hadrons
in $\sqrt{s_{NN}}$=200\,AGeV Au-Au Collisions}

\author{Min He}
\affiliation{Cyclotron Institute and Department of Physics \& Astronomy,
Texas A\&M University, College Station, TX 77843-3366, USA}

\author{Rainer J.\ Fries}
\affiliation{Cyclotron Institute and Department of Physics \& Astronomy,
  Texas A\&M University, College Station, TX 77843-3366, USA}
\affiliation{RIKEN/BNL Research Center, Brookhaven National Laboratory,
Upton, NY 11973, USA}

\author{Ralf Rapp}
\affiliation{Cyclotron Institute and Department of Physics \& Astronomy,
Texas A\&M University, College Station, TX 77843-3366, USA}

\date{\today}

\begin{abstract}
We revisit the use of ideal hydrodynamics to describe bulk- and
multistrange-hadron observables in nuclear collisions at the
Relativistic Heavy Ion Collider. Toward this end we augment the
2+1-dimensional code ``AZHYDRO" by employing (a) an equation of
state based on recent lattice-QCD computations matched to a
hadron-resonance gas with chemical decoupling at $T_{\rm
ch}\simeq$160\,MeV, (b) a compact initial density profile, (c) an
initial-flow field including azimuthal anisotropies, and (d) a
sequential kinetic decoupling of bulk ($\pi$, $K$, $p$) and
multistrange ($\phi$, $\Xi$, $\Omega$) hadrons at $T\simeq110$\,MeV
and 160\,MeV, respectively. We find that this scheme allows for a
consistent description of the observed chemistry,
transverse-momentum spectra and elliptic flow of light and strange
hadrons.
\end{abstract}


\maketitle

%%%%%%%%%%%%%%%%%%%%%%%%%%%%%%%%%%%%%%%%%%%%%%%%%%%%%%%%
\section{Introduction}
\label{sec_intro}
%%%%%%%%%%%%%%%%%%%%%%%%%%%%%%%%%%%%%%%%%%%%%%%%%%%%%%%
Experiments at the Relativistic Heavy Ion Collider (RHIC) and the
Large Hadron Collider (LHC)~\cite{Adams:2005dq,Aamodt:2010pa}
suggest that a quark gluon plasma (QGP) is created in
ultra-relativistic heavy-ion collisions (URHICs) which is strongly
coupled and behaves like a near-perfect liquid\-
~\cite{Gyulassy:2004zy,Shuryak:2008eq}. In particular, the use of
ideal relativistic hydrodynamics enabled a good description of
several bulk hadron-observables encompassing more than 90\% of the
produced
particles~\cite{Kolb:2003dz,Teaney:2001av,Hirano:2002ds,Kolb:2002ve,Huovinen:2006jp,Nonaka:2006yn,Hama:2005dz,Niemi:2008ta,Schenke:2010nt,Fries:2010ht}.
A rapid thermalization of the medium, leading to collective
phenomena including elliptic flow, could be established and are key
to our understanding of the macroscopic properties of the fireball.
The success of ideal hydrodynamics and the conclusion that
dissipative effects appear to be small has more recently led to
efforts to quantify these by employing second-order viscous
hydrodynamics~\cite{Romatschke:2007mq,Song:2007ux,Dusling:2007gi,Heinz:2009xj,Song:2010mg,Schenke:2010rr,Niemi:2011ix,Roy:2011xt,Shen:2011eg}.
This goal clearly requires a good control over any remaining
uncertainties within the hydrodynamic framework, e.g.,  the equation
of state (EoS), initial conditions and implementations of the
freezeout scenario and/or final-state transport.
%In particular, we would like to measure how close the shear
%viscosity-to-entropy ratio $\eta/s$ is to the
%postulated universal bound of $1/4\pi$ \cite{Policastro:2001yc}.

Some of these aspects and their interplay, common to both ideal and
viscous hydrodynamics, are not well understood to date. For example,
the elliptic flow, $v_2$, calculated in ideal hydrodynamics seemed
to favor an EoS with a strong first order phase
transition~\cite{Huovinen:2005gy}, contradicting the finite-$T$
cross-over transition now firmly established in lattice quantum
chromodynamics (QCD)~\cite{Aoki:2006we,Cheng:2009zi}. On the other
hand, the recent progress in solving the so-called Hanbury
Brown-Twiss (HBT) puzzle required several effects to increase the
transverse expansion, including viscosities, initial flow and a hard
EoS without phase transition~\cite{Pratt:2008qv}.
%$R_{\mathrm{out}}/R_{\mathrm{side}}$
The development of initial flow, prior to the thermalization time
assumed in hydrodynamics, can be expected on rather general
grounds~\cite{Fries:2005yc,Vredevoogd:2008id}, but has only been
studied in few works to date~\cite{Kolb:2002ve,Heinz:2002rs,
Krasnitz:2002ng,Broniowski:2008qk,Pratt:2008qv}. The initial density
profile and initial fluctuations are not yet well constrained from
first principles, with both Glauber- and Color-Glass Condensate
(CGC)-based approaches currently being
investigated~\cite{Hirano:2005xf,Qiu:2011hf}. If dissipative effects
become large, a transition to a transport treatment of the bulk is
in order, which has been studied by coupling hadronic cascades to
hydrodynamic evolutions of the
QGP~\cite{Bass:2000ib,Hirano:2005xf,Nonaka:2006yn,Song:2010mg}.
However, it is quite possible that the viscosity in the hadronic
phase remains small for a significant range of temperatures below
$T_c$, especially if partial chemical equilibrium is implemented.
The latter becomes problematic in cascade models if the inverse of
reactions with multi-particle final states need to be accounted for,
as, e.g., for baryon-antibaryon annihilation into
mesons~\cite{Rapp:2000gy,Greiner:2000tu}.
%The concept of separate chemical and kinetic freezeouts has been
%first introduced in Ref.~\cite{Bebie:1991ij} based on the realization
%that elastic hadronic cross sections (usually proceeding through an
%intermediate resonance) are much larger than inelastic ones.
To date, sequential chemical and thermal freezeouts in URHICs are
experimentally well established, signified by statistical-model fits
to hadron
abundances~\cite{BraunMunzinger:2003zd,Becattini:2000jw,Abelev:2008ez}
on the one hand (yielding $T_{\rm ch}\approx 160~{\rm MeV}$), and
empirical blast-wave fits to transverse-momentum ($p_T$) spectra of
bulk hadrons ($\pi$, $K$, $p$)~\cite{Abelev:2008ez} on the other
hand (yielding $T_{\rm fo}\approx 100~{\rm MeV}$). In the hadronic
EoS figuring into hydrodynamics the number conservation of stable
hadrons ($\pi$, $K$, $p$, $\bar p$, $\eta$, etc.) between $T_{\rm
ch}$ and $T_{\rm fo}$ can be enforced by introducing pertinent
chemical potentials~\cite{Bebie:1991ij}. This has been implemented
into ideal hydrodynamic
models~\cite{Hirano:2002ds,Teaney:2002aj,Kolb:2002ve,Huovinen:2007xh}.
It was found that bulk-hadron $p_T$ spectra can be reproduced well
at $T_{\rm fo}$, but the previous agreement with the observed
elliptic flow, $v_2(p_T)$,
deteriorates~\cite{Kolb:2002ve,Hirano:2005wx,Huovinen:2007xh}, i.e.,
it is overpredicted. In viscous hydrodynamics, a systematic
investigation of the effects of chemical freezeout on bulk
observables are still in their
beginnings~\cite{Huovinen:2009yb,Niemi:2011ix,Shen:2010uy}.

The spectra and $v_2$ of multistrange particles have received
relatively little attention in hydrodynamic calculations thus far.
The $\phi$, $\Xi$ and $\Omega$ have no well established resonances
with bulk hadrons, and elastic $t$-channel exchange processes are
suppressed by the Okubo-Zweig-Iizuka (OZI) rule. Therefore,
multistrange hadrons are not expected to undergo significant
rescattering in the hadronic phase and should decouple from the
system early~\cite{Shor:1984ui,vanHecke:1998yu}. This is supported
by experimental data from
RHIC~\cite{Adams:2005dq,Xu:2007zzb,Abelev:2008fd}. Compared to bulk
hadrons multistrange particles thus reflect more directly the
collective dynamics of the partonic stage of the fireball and can
provide a significant but often neglected constraint on hydrodynamic
evolution models. For example, in the 2+1-dimensional
hydro-simulations of Ref.~\cite{Kolb:2003dz}, $\Omega^-$ freezeout
has to be carried well into the hadronic phase to be compatible with
the experimental $p_T$ spectra.

In this work we will revisit to what extent \emph{ideal}
hydrodynamics is capable of providing a realistic description of the
bulk evolution of the medium in Au-Au collisions at RHIC. With
``realistic" we mean, on the one hand, a consistent description of
light- and strange-hadron observables encompassing their abundances,
$p_T$-spectra and $v_2(p_T)$ at midrapidity for semi-/central
collisions.
%Even though the number of constraints is larger than usual (by
%including multi-strange hadrons) the consistency with data is rather good.
On the other hand, we also refer to inputs to, and assumptions in,
the hydrodynamic treatment which are within the uncertainties
described above. This includes (i) a state-of-the-art equation of
state adopted from lattice QCD in the QGP phase, matched to a hadron
resonance gas with hadrochemical freezeout; (ii) a sequential
freezeout of bulk-hadron chemistry and kinetics, as well as
simultaneous kinetic and chemical freezeout of multistrange hadrons
(at the universal chemical freezeout); (iii) a compact initial
density profile with relatively large gradients and ``reasonable"
initial radial and elliptic flow.
%Note that both of these conditions are also part of the solution to the
%HBT puzzle \cite{Pratt:2008qv}.
We will also elaborate on arguments why the viscosity in the
hadronic phase of URHICs could be rather small. We will, however,
neglect the role of initial-state fluctuations, which are presumably
the main source of higher flow harmonics, but do not seem to affect
$v_{0,2}$ much~\cite{Schenke:2010rr}.

Our computations build on the existing and readily available ideal
hydrodynamic code package AZHYDRO developed by Kolb and
Heinz~\cite{Kolb:2003dz}. One of our goals in performing a tune of
AZHYDRO is to provide an easily accessible yet realistic background
medium for quantitative applications to electromagnetic and
heavy-flavor probes. In particular, we do not imply to supersede
emerging viscous hydrodynamic calculations, which, once fully
developed, are expected to become the tool of choice.

Our article is organized as follows. In Sec.~\ref{sec_amend} we
briefly recall the basic features of the 2+1 dimensional hydro-code
AZHYDRO and then focus on its amendments as applied in the present
work. In Sec.~\ref{sec_bulk}, we analyze the qualitative impact of
our amendments on the evolution of the bulk medium, i.e., its radial
and elliptic flow, and potential ramifications for hadron spectra,
especially freezeout properties.  In Sec.~\ref{sec_fits}, we conduct
quantitative fits to bulk- and multistrange-hadron observables,
encompassing multiplicities, $p_T$ spectra and elliptic flow. We
summarize and conclude in Sec.~\ref{summary}.

%%%%%%%%%%%%%%%%%%%%%%%%%%%%%%%%%%%%%
\section{AZHYDRO and its Amendments}
\label{sec_amend}
%%%%%%%%%%%%%%%%%%%%%%%%%%%%%%%%%%%%
The starting point for ideal hydrodynamics (IH) are the equations
for the conservation of energy and momentum,
\begin{equation}
  \partial_\mu T^{\mu\nu} =0 \ ,
\end{equation}
formulated in terms of the energy-momentum tensor, $T^{\mu\nu}$. As
usual, the latter is given by the energy density, $e$, and pressure,
$p$, in the local rest frame assuming kinetic equilibrium, together
with a flow field $\mathbf{v}$ describing the collective motion of
the fluid cells. Other conserved currents (e.g., for baryon number),
can be introduced as appropriate, see, e.g.,
Refs.~\cite{Kolb:2003dz,Huovinen:2006jp} for reviews. The system has
to be closed by specifying an equation of state $p(e)$, initialized
at a thermalization time $\tau_0$ (typically with an initial entropy
density and flow field) and frozen out in the dilute stage
(typically at a final energy density using the Cooper-Frye
prescription~\cite{Cooper:1974mv} to convert the fluid cells into
hadron spectra). In AZHYDRO~\cite{Kolb:2002ve,Kolb:2003dz} this is
done in a longitudinally boost-invariant setup leading to a 2+1
dimensional evolution. In the following we discuss in more detail
our modifications to two of the above ingredients, namely the
equation of state (Sec.~\ref{ssec_eos}) and the initial conditions
(Sec.~\ref{ssec_ini}).

%\subsection{Partial Chemical Equilibrium and Sequential FreezeOut}

%Given the fact that
%in an early ideal hydro implementation~\cite{Kolb:2003dz}, $\Omega$
%was taken to freeze out together with bulk particles, it seems more
%worthy and necessary to systematically explore the freezeout
%dynamics of multistrange particles and bulk particles separately in
%hydrodynamics. The present work serves to fill this hole in
%literature.

% All these ingredients (a stiffer EoS, more compact
%initial density profile and pre-equilibrium flow) point to
%increasing the radial flow in the early partonic stage and help
%establish the earlier freezeout of multistrange particles than bulk
%particles. These ingredients have also been found crucial in
%resolving the HBT puzzle~\cite{Pratt:2008qv}. Shear viscosity to the
%entropy ratio ($\eta/s$) in high temperature phase has been found to
%be very small~\cite{Song:2010mg}. In low temperature hadronic phase,
%rho-meson thermal with is found to be large ($\sim 200~{\rm MeV}$)
%even at as low as $T=100~{\rm MeV}$~\cite{Rapp:2009yu}, signalling
%strongly coupled behavior of a pion liquid whose $\eta/s$ may not be
%that large. In the present work, we leave the viscous effects and
%attempt to see if one can describe bulk as well as multistrange
%particles' observables in a sequential kinetic freezeout scenario
%with ideal hydro.

%%%%%%%%%%%%%%%%%%%%%%%%%%%%%%%%%%%%%%%%%%%%%%%%%%%%%%%%%%%%%%%%
\subsection{Equation of State}
\label{ssec_eos}
\begin{figure}[!t]
\includegraphics[width=\columnwidth]{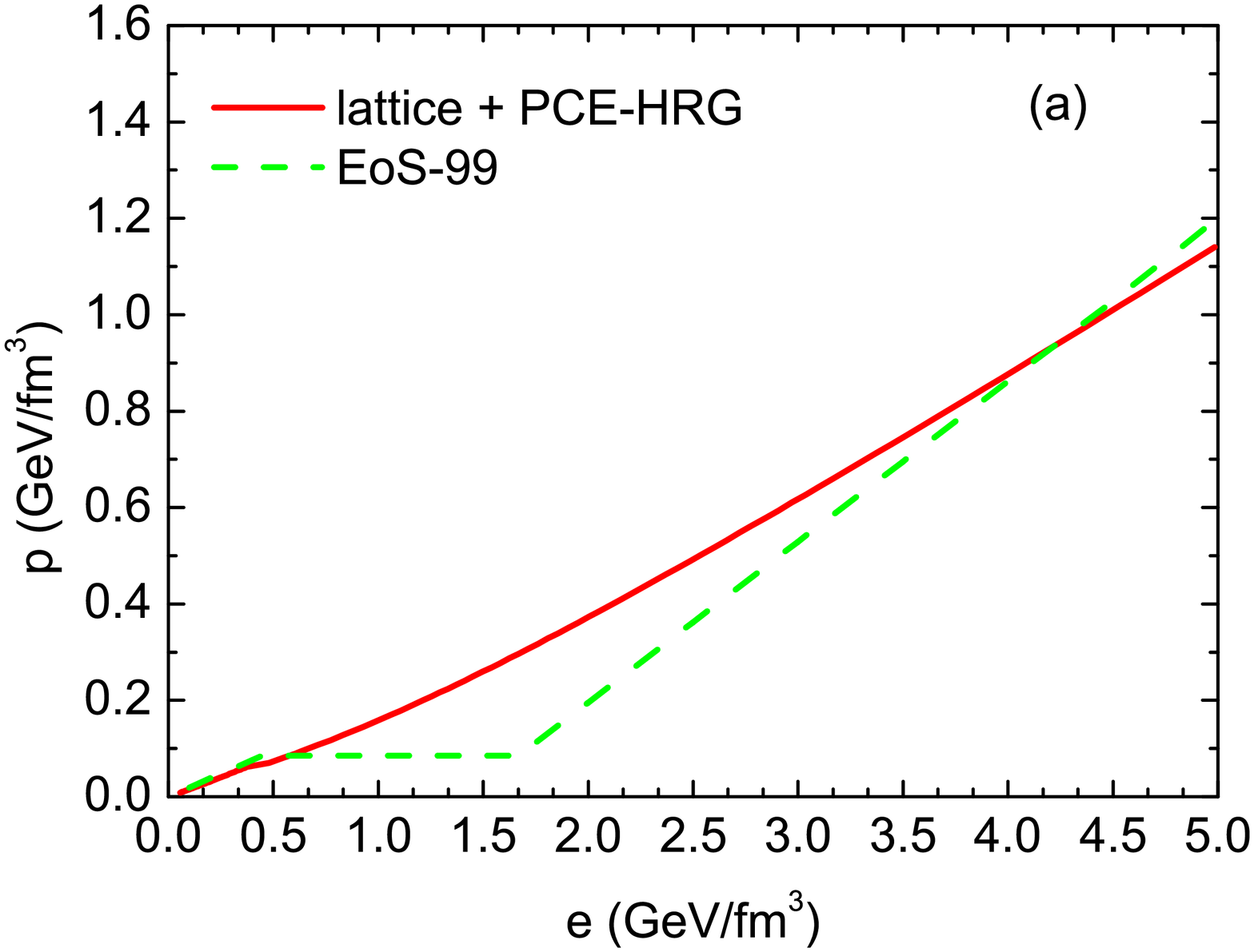}
\includegraphics[width=\columnwidth]{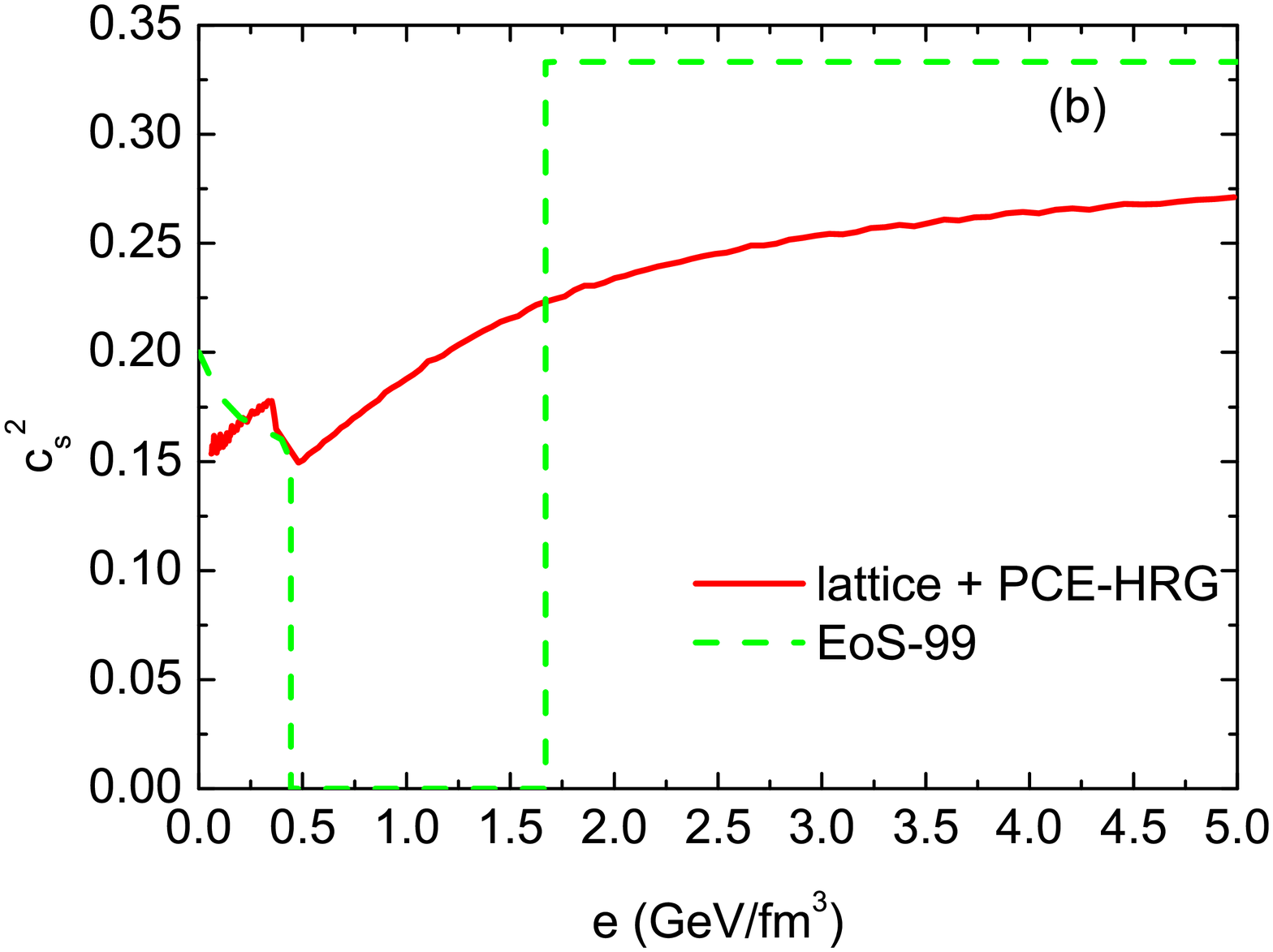}
\caption{(Color online) Comparison of pressure (upper panel) and speed
of sound squared (lower panel) vs.~energy density as obtained from
EoS-99 (dashed lines) and our new equation of state (solid lines).}
\label{EOS}
\end{figure}

%%%%%%%%%%%%%%%%%%%%%%%%%%%%%%%%%%%%%%%%%%%%%%%%%%%%%%%%%%%%%%%%
The default AZHYDRO code (version
v0.2)~\cite{Kolb:2002ve,Kolb:2003dz} employs an equation of state
(labeled ``EoS-99") consisting of an ideal massless quark-gluon gas
at high temperatures and a hadron resonance gas (HRG) in partial
chemical equilibrium at low temperatures\footnote{Recently, a patch
was released for AZHYDRO by Molnar and Huovinen wich also contains a
lattice-based EoS~\cite{Huovinen:2009yb}.}. The two parts are
matched via a Maxwell construction at a critical temperature
$T_c=165~{\rm MeV}$ in a first-order phase transition with a mixed
phase, see dashed line in the upper panel of Fig.~\ref{EOS}. The
sound velocity $c_s$ vanishes in the mixed phase (see the lower
panel of Fig.~\ref{EOS}), resulting in a vanishing acceleration over
a relatively long duration. Consequently, the radial flow of the
system largely stalls until the end of the mixed phase. In practice,
this implies, e.g., that multistrange particles like the $\Omega^-$
baryon have to be decoupled close to the kinetic freezeout of the
bulk particles, at $T_{\rm fo}\simeq100$~MeV, to reproduce the
pertinent experimental spectra at RHIC~\cite{Kolb:2003dz}.

Recent lattice-QCD (lQCD) calculations have now established that, at
vanishing baryon chemical potential ($\mu_B=0$), the transition
between the QGP and hadronic matter is a smooth
crossover~\cite{Aoki:2006we,Cheng:2009zi}. The pseudo-critical
deconfinement transition temperature obtained by the two leading
lQCD groups~\cite{Cheng:2009zi,Borsanyi:2010bp} has now converged to
a common value of around $T_c\simeq170~{\rm
MeV}$~\cite{Aoki:2009sc}. Some differences remain regarding the
interaction measure, $I=e-3p$~\cite{Borsanyi:2010cj,Soldner:2010xk}.
The calculations by the Wuppertal-Budapest (WB)
collaboration~\cite{Borsanyi:2010cj} reproduce the HRG results in
chemical equilibrium below $T_c$, while their lattice action is not
ideally suited to the high-$T$ limit where their thermodynamic
quantities come out slightly below the results of the HotQCD
collaboration~\cite{Soldner:2010xk}. In our fit we decided to employ
the WB results in the transition region and to smoothly match on to
the HotQCD results for $T\gsim 180$~MeV. Once $I(T)$ is specified we
calculate the EoS, $p(e)$, following the procedure described in
Ref.~\cite{Borsanyi:2010cj}. The resulting EoS describes strongly
interacting matter in thermal and chemical equilibrium. However, in
URHICs it is well-known that hadron ratios freeze out at a
temperature of $T_{\rm
ch}\simeq160$\,MeV~\cite{BraunMunzinger:2003zd,Becattini:2000jw,Abelev:2008ez}.
To account for the departure from chemical equilibrium in the
hadronic phase below this temperature, we follow the approach in
Ref.~\cite{Rapp:2002fc}, by introducing effective chemical
potentials for hadrons which are stable under strong interactions,
i.e., pions, kaons, etas, nucleons and antinucleons, including their
feeddown contributions (e.g., $\rho\to2\pi$ with
$\mu_\rho=2\mu_{\pi}$), usually referred to as ``partial chemical
equilibrium (PCE). The conservation of antibaryon number is of
particular importance since, in turn, it triggers the build-up of
pion chemical potentials~\cite{Rapp:2002fc}. We have verified that
the pertinent chemical-equilibrium EoS is consistent with the lQCD
results. In the construction of our URHIC EoS we then replace the
chemical-equilibrium part with the PCE part. We will refer to the
resulting EoS as ``latPHG" EoS\footnote{Following the systematic
analysis of hadron observables in Ref.~\cite{Abelev:2008ez} we also
introduce a ``strangeness suppression" factor $\gamma_s=\gamma_{\bar
s}\simeq0.85$ for each net (anti-) strange quark in a hadron.},
which is compared to EoS-99 in Fig.~\ref{EOS}. As expected, the most
notable differences are in the transition regime, where the pressure
in the latPHG EoS is enhanced; but, most importantly, the speed of
sound remains large, $c_s^2 = \partial p/\partial e\simeq0.15-0.20$,
compared to zero in EoS-99. As is well known, this will have
significant ramifications for the radial flow, especially toward the
end of the transition, $e\simeq0.5$~GeV/fm$^3$, and in connection
with the kinetic freezeout of multistrange particles.

%%%%%%%%%%%%%%%%%%%%%%%%%%%%%%%%%%%%%%%%%%%%%%%%%%%%%%%%%%%%%%
\subsection{Initial Conditions}
\label{ssec_ini}
%%%%%%%%%%%%%%%%%%%%%%%%%%%%%%%%%%%%%%%%%%%%%%%%%%%%%%%%%%%%%%

Initial conditions currently constitute one of the largest
uncertainties in hydrodynamic simulations of
URHICs~\cite{Song:2010mg}. In default AZHYDRO, a combination of
wounded-nucleon and binary-collision density is used to initialize
the entropy density,
\begin{equation}
s(\tau_0,x,y;b)={\rm const} \ [0.25  \frac{n_{\rm BC}(x,y;b)}{n_{\rm
BC}(0,0;0)}+0.75 \frac{n_{\rm WN}(x,y;b)}{n_{\rm WN}(0,0;0)}] \ ,
\end{equation}
at an initial time of $\tau_0=0.6$ fm/$c$ for each impact parameter
$b$. This ansatz leads to good agreement with the observed
centrality dependence of charged particle
multiplicities~\cite{Kolb:2003dz,Kolb:2001qz}. Other initial density
profiles, e.g., inspired by the CGC, have also been
used~\cite{Kharzeev:2002ei,Drescher:2006ca}.

However, to accelerate the build-up of radial flow, which is
essential for describing spectra of multistrange particles at
$T_{\rm ch}$, a compact initial profile with large initial pressure
gradients is favored. Such a profile is furthermore an essential
ingredient to a realistic description of HBT
radii~\cite{Pratt:2008qv}. As a limiting case, we choose the
entropy-density profile to be solely proportional to the
binary-collision density
\begin{equation}
\label{snBC}
s(\tau_0,x,y;b) = C(b) \ n_{\rm BC}(x,y;b) \, .
\end{equation}
While our fits below favor compact profiles they do not necessarily
dictate collision scaling, albeit it turns out to work well. We
calculate $n_{\rm BC}$ using an optical Glauber
model~\cite{Kolb:2003dz,Kolb:2001qz}. Compact profiles have also
been used in some other hydrodynamic
simulations~\cite{Huovinen:2007xh,Niemi:2011ix,Broniowski:2008vp}.
Since the particle multiplicity does not scale with $n_{\rm BC}$,
the coefficient in Eq.~(\ref{snBC}) needs to become
$b$-dependent~\cite{Huovinen:2007xh}.

Another uncertainty in the initial conditions concerns the
possibility of pre-equilibrium flow. In most hydrodynamic
simulations, the initial transverse collective velocity is assumed
to be zero. However, it has been argued that flow can easily emerge
before kinetic equilibrium is
established~\cite{Fries:2005yc,Vredevoogd:2008id}. An initial radial
flow field has been implemented into a few calculations and shown to
improve the agreement with bulk-hadron
data~\cite{Kolb:2002ve,Broniowski:2008qk} (e.g., by reducing the
final $v_2$ in calculations with PCE in the hadronic EoS, or by
improving on the HBT data). In the present work, we go one step
further and adopt a non-trivial pre-equilibrium flow field including
finite ellipticity. Specifically, we employ the empirical ansatz
proposed in Ref.~\cite{Retiere:2003kf}, which was successfully used
to fit bulk and multistrange observables in a sequential kinetic
freezeout scenario~\cite{He:2010vw}. The transverse velocity is
parameterized in terms of the spatial coordinates $r$ and $\phi_s$
as
\begin{equation}
\label{vTxy}
v(r,\phi_s)=\widetilde{r}[\alpha_0 + \alpha_2{\rm cos}(2\phi_b)] \ ,
\end{equation}
where $\alpha_0$ quantifies the surface radial flow and $\alpha_2$
the anisotropy. The azimuthal angle $\phi_b$ of the flow vector can
be tilted away from $\phi_s$ through
\begin{equation}
\label{phibphis}
{\rm tan}\phi_b=\kappa \frac{R_{\rm x}^2}{R_{\rm y}^2}{\rm
tan}\phi_s \, ,
\end{equation}
where $\kappa$ is a tunable parameter dependent on the impact
parameter $b$, and $R_{\rm x}= R_0-b/2$, $R_{\rm
y}=\sqrt{R_0^2-(b/2)^2}$ are the short and long half-axis of the
initial ellipse, and $R_0$ is the radius of the colliding nuclei
(with identical $A$). Finally, the normalized radius $\widetilde{r}$
in Eq.~(\ref{vTxy}) is
\begin{equation}
\label{rtild}
\widetilde{r}=\sqrt{\frac{r^2}{R_{\rm x}^2} \cos^2\phi_s
              + \frac{r^2}{R_{\rm y}^2} \sin^2 \phi_s} \ .
\end{equation}
We keep the initial time $\tau_0 =0.6$\,fm/$c$ as in the default AZHYDRO.

%%%%%%%%%%%%%%%%%%%%%%%%%%%%%%%%%%%%%%%%%%%%%%%%%%%%%%%%%%%%%%
\section{Bulk Evolution}
\label{sec_bulk}
%%%%%%%%%%%%%%%%%%%%%%%%%%%%%%%%%%%%%%%%%%%%%%%%%%%%%%%%%%%%%%
\begin{figure}[!t]
\includegraphics[width=1.0\columnwidth]{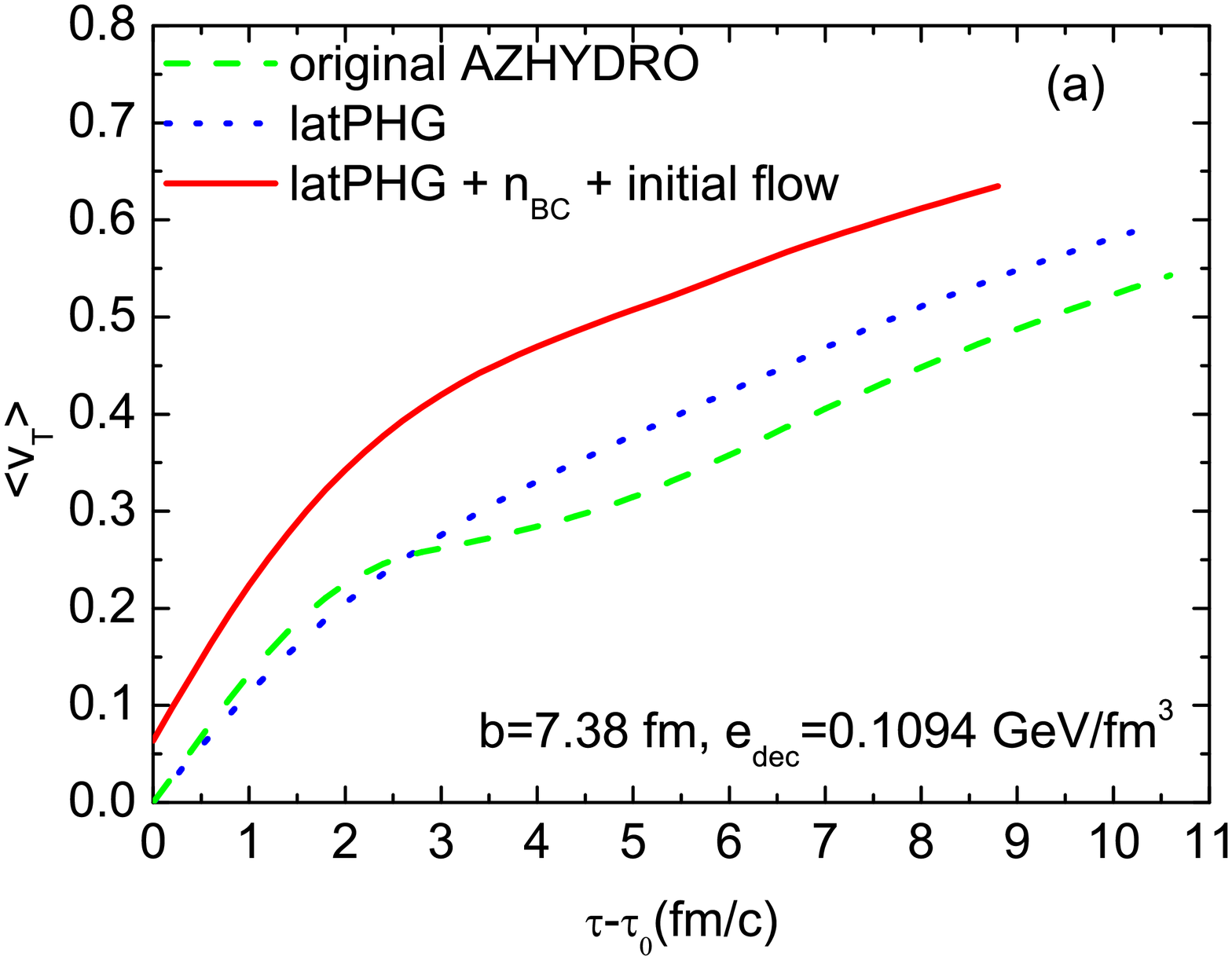}
\includegraphics[width=1.0\columnwidth]{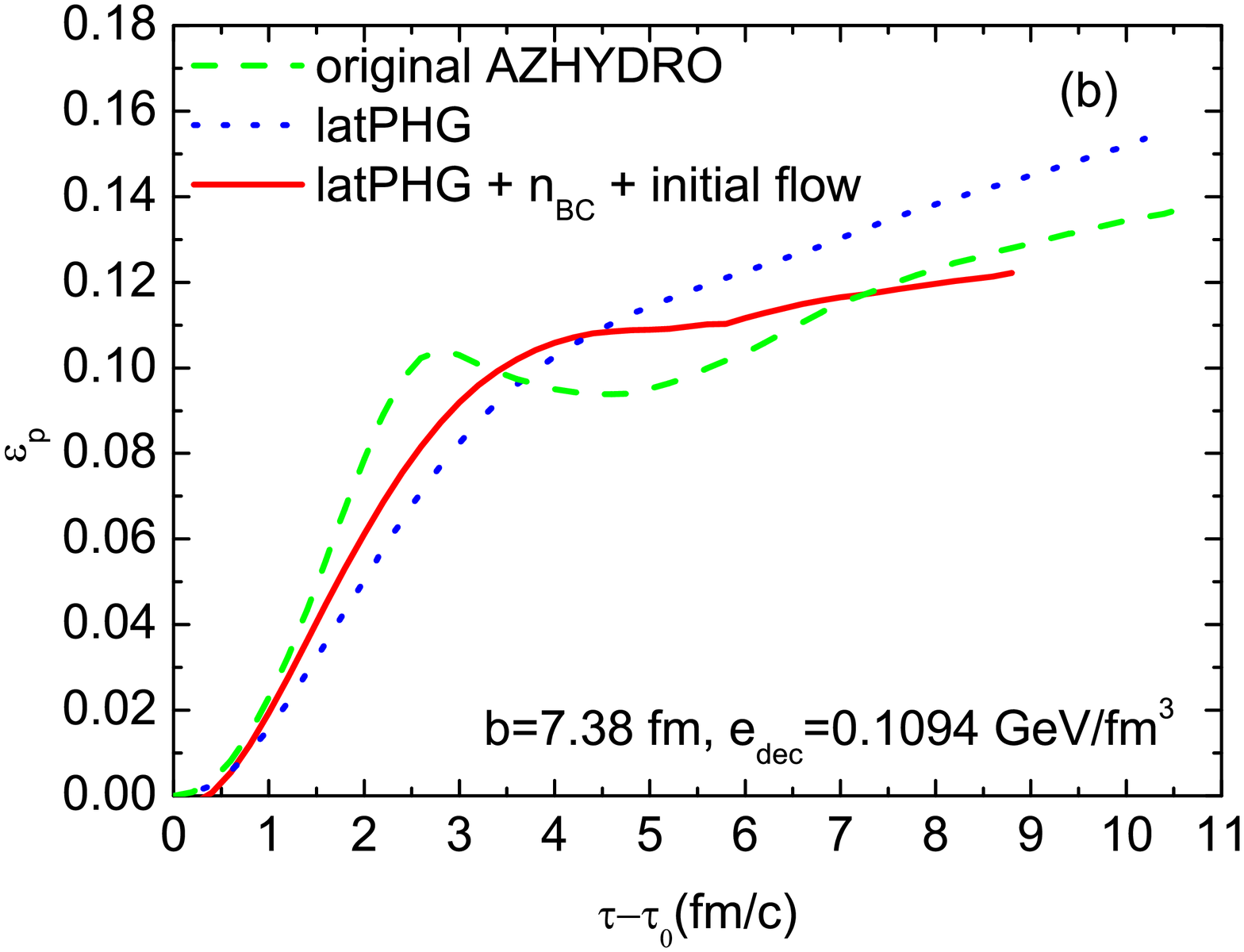}
\includegraphics[width=1.0\columnwidth]{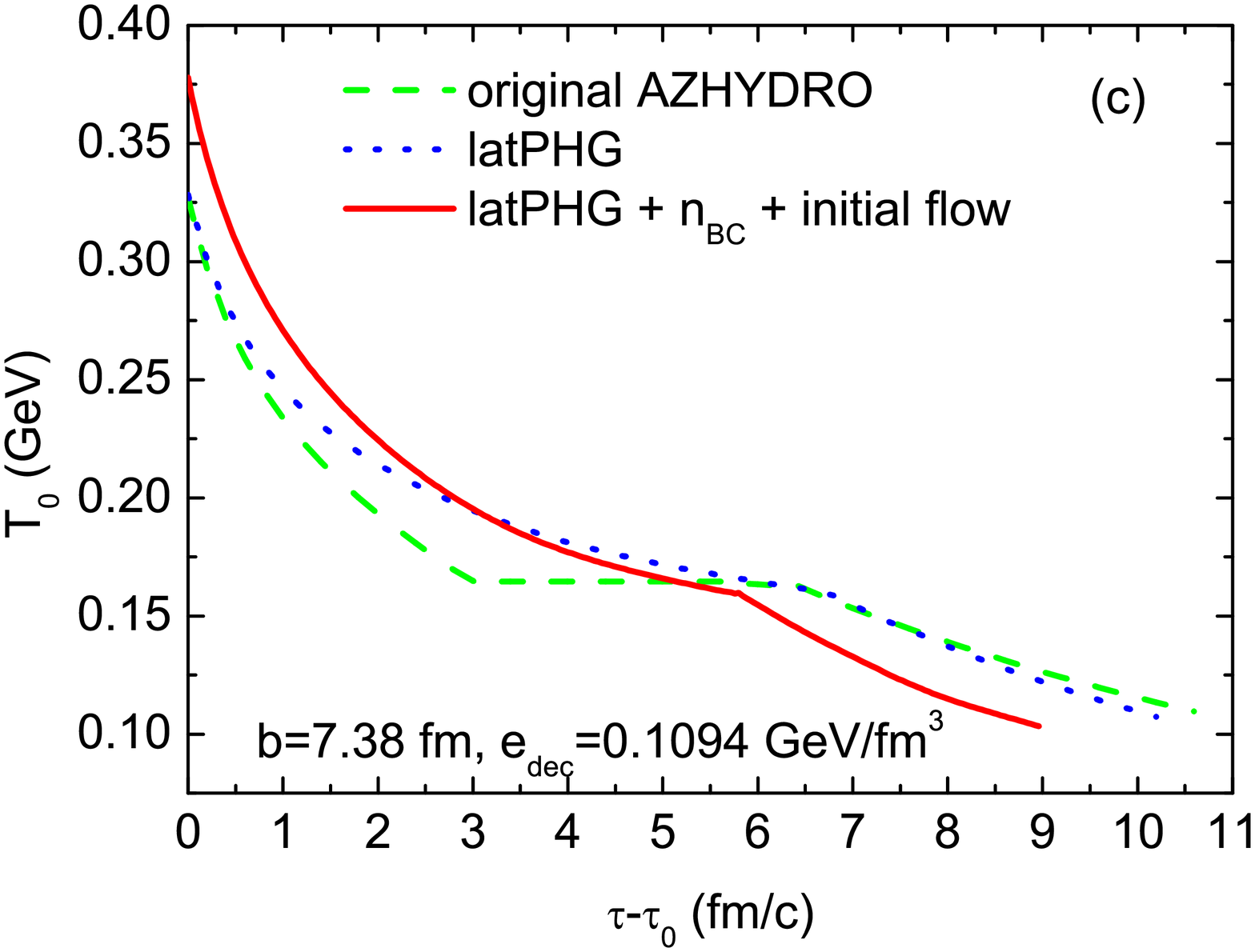}
\caption{(Color online) (a) Proper-time evolution of the average
transverse radial velocity, $\langle v_T\rangle$, for different
hydro scenarios: (i) default AZHYDRO with EoS-99 (green dashed
curve), (ii) default AZHYDRO with the new EOS (blue dotted curve),
and (iii) the new EoS with initial flow and compact initial density
profile (red solid curve). (b) The same comparison for the time
evolution of the energy-momentum anisotropy, $\varepsilon_P$. (c)
The evolution of the temperature in the central cell of the
transverse plane.} \label{bulk_evolution}
\end{figure}
Before turning to quantitative fits to hadron spectra in the
following section, we first illustrate the effects of our amendments
on the bulk-matter evolution of a fireball in semicentral Au-Au at
RHIC. The bulk evolution in a hydrodynamic system can be
characterized by the time evolution of the average transverse radial
flow, $\langle v_T \rangle$, and of the anisotropy of the
energy-momentum tensor~\cite{Kolb:2003dz},
\begin{equation}
\varepsilon_P =\frac{\int dxdy(T^{\rm xx}-T^{\rm yy})}{\int
dxdy(T^{\rm xx}+T^{\rm yy})} \ .
\end{equation}
In Fig.~\ref{bulk_evolution}, we compare the longitudinal
proper-time evolution of both quantities in default AZHYDRO to
simulations with updated EoS and with both updated EoS and modified
initial conditions. The parameters for the latter correspond to the
final data fit discussed in Sec.~\ref{ssec_pt}. Since the new EoS
does not feature a vanishing acceleration at (or around) $T_c$, the
knee in $\langle v_T(\tau) \rangle$ and the dip in
$\varepsilon_P(\tau)$ disappear. However, the acceleration during
the first 3\,fm/$c$ is slightly lagging behind in latPHG relative to
EoS99 due to the non-ideal behavior of the former. As a result, the total
radial flow is not very different at the beginning of the hadronic
phase, which also implies that the anisotropy keeps increasing
significantly in the hadronic evolution of both scenarios. These
(possibly undesired) properties can be modified by introducing more
compact initial profiles and initial flow, as shown by the pertinent
solid lines in Fig.~\ref{bulk_evolution}: the radial flow increases
by ca.~50\% at $T_c$, while the eccentricity essentially levels off
thereafter. Both features are crucial in fitting multistrange
particle spectra at $T_c$, and also improve the description of bulk
particles at $T_{\rm fo}$ (in particular the reduced anisotropy, but
the increased final flow also helps to describe the $p_T$ spectra
out to higher $p_T$ than before). The more rapid expansion of the
fireball shortens the lifetime of the fireball by almost 15\%, from
$\sim$10.6\,fm/$c$ in default AZHYDRO to $\sim$9.0\,fm/$c$ in the
fully amended case (both with $e_{\rm{dec}}=0.1094~{\rm GeV/fm^3}$).

The lower panel in Fig.~\ref{bulk_evolution} shows the time
evolution of the temperature in the central cell of the transverse
plane. While the initial temperature in the fully amended case is
higher than in the default AZHYDRO (due to more compact
initial-entropy profile), it turns out that the duration of the
hadronic phase in the two cases is comparable (note that
$T_c=170$\,MeV vs $165$\,MeV in the amended and default AZHYDRO,
respectively).

%%%%%%%%%%%%%%%%%%%%%%%%%%%%%%%%%%%%%%%%%%%%%%%%%%%%%%%%%%%%%%
\section{Hadron Observables}
\label{sec_fits}
%%%%%%%%%%%%%%%%%%%%%%%%%%%%%%%%%%%%%%%%%%%%%%%%%%%%%%%%%%%%%%
The amended AZHYDRO as described above has been utilized to conduct
``eye-ball" fits to simultaneously describe spectra and $v_2$ of
bulk and multistrange hadron (at $T_{\rm fo}$ and $T_{\rm ch}$,
respectively) in Au-Au collisions at full RHIC energy. We have
focused on two centralities, 0-5\% (``central") and 20-30\%
(``semicentral"), which, for simplicity, we have approximated by
fixed impact parameters ($b$=2.3 and 7.38\,fm, respectively)
corresponding to participant numbers,  $N_{\rm part}$, calculated in
the optical Glauber model~\cite{Abelev:2008ez} for the two
experimental selections. The central initial-entropy density,
$s_0\equiv s(\tau_0,0,0;b)$, is then adjusted to the experimental
hadron multiplicities for the two centralities, evaluated at kinetic
freezeout in the evolution ($e_{\rm fo}=0.1094~{\rm GeV/fm^3}$, or
$T_{\rm fo}=110$\,MeV) incorporating resonance decays (the spectra
and feeddown from multistrange hadrons, for which we assume early
kinetic freezeout, are evaluated at $T_{\rm ch}=160$\,MeV). The two
parameters in the initial-flow field basically control the radial
flow ($\alpha_0$) and, subsequently, the elliptic flow ($\alpha_2$).
Within our accuracy they turn out centrality-independent, provided
the parameter for the ``tilt" between position vector and flow
vector is suitably increased for more peripheral collisions (which
is consistent with intuition, e.g., going to one in the academic
limit of $b$=0). The resulting parameter values are summarized in
Tab.~\ref{init_parameters}. The kinetic freezeout temperature,
$T_{\rm fo}=110$\,MeV, also turns out to be identical for the two
centrality classes within our accuracy.
\begin{table}
\begin{tabular}{lcccccc}
\hline\noalign{\smallskip}
 $~~~$ & $s_0~({\rm fm^{-3}})$ & $T_0~({\rm MeV})$ & $\alpha_0$ & $\alpha_2$ & $\kappa$ \\
\noalign{\smallskip}\hline\noalign{\smallskip}

0-5\%       & 159.5 & 399.9 & 0.13  &  0.004  &  1.0  \\
20-30\%     & 133.1 & 377.9 & 0.13  &  0.004  &  4.2  \\

\noalign{\smallskip}\hline
\end{tabular}
\caption{Initialization parameters for central (0-5\%, $b=2.3~{\rm
fm}$) and semicentral (20-30\%, $b=7.38~{\rm fm}$) Au-Au
($\sqrt{s_{NN}}$=200\,GeV) collisions; $s_0$ is the initial-entropy
density in the center of the transverse plane and $T_0$ is the
pertinent temperature resulting from a collision-density overlap;
$\alpha_0$, $\alpha_2$ and $\kappa$ parameterize the initial
anisotropic flow profile, cf.~ Sec.~\ref{ssec_ini}.}
\label{init_parameters}
\end{table}

A few remarks are in order concerning the applicability of
hydrodynamics in the (later stages of the) hadronic phase. It is
often argued that the latter carries large viscosity thus mandating
a transport treatment. However, there are several arguments
suggesting that the hadronic evolution in URHICs does not carry
large $\eta/s$ ratios. First, since the QCD transition at $\mu_B$=0
is presumably close to a second order one, it is likely that
$\eta/s$ possesses a minimum around the pseudocritical temperature
of $T_c\simeq170$\,MeV. The question then is how fast $\eta/s$
increases with decreasing
$T$~\cite{Prakash:1993bt,Csernai:2006zz,Chen:2007xe,Chakraborty:2010fr}.
Dilepton measurements (and pertinent calculations) at the SPS show
that the average $\rho$-meson width in the hadronic phase is
substantially broadened, by more than 200\,MeV
(cf.~Ref.~\cite{Rapp:2009yu} for a recent review). This corresponds
to a mean-free-path of below 1\,fm, and is comparable to the thermal
kinetic energy of the $\rho$, $KE\simeq 1.5T\simeq 225$\,MeV. Even
at thermal freezeout the $\rho$ is still broadened by ca.~100\,MeV.
From another angle, but using similar techniques (\ie, effective
hadronic interactions), charm diffusion has been evaluated in
hadronic matter in Ref.~\cite{He:2011yi}. While the diffusion
coefficient (which, in units of the thermal wavelength, $1/(2\pi
T)$, is roughly proportional to $\eta/s$) increases appreciably
toward lower temperatures in equilibrium matter, the inclusion of
effective chemical potentials only leads to a $\sim$30\% increase
when going down from $T$=170\,MeV to 100\,MeV. We thus believe that
viscosity effects in hadronic matter as formed in URHICs may be
significantly smaller than commonly assumed.

In the following sections we present the results for the
observables, starting with inclusive yields (hadro-chemistry,
Sec.~\ref{ssec_chem}) and then becoming more differential with $p_t$
spectra (Sec.~\ref{ssec_pt}) and azimuthal dependencies
(Sec.~\ref{ssec_v2}).

%%%%%%%%%%%%%%%%%%%%%%%%%%%%%%%%%%%%%%%%%%%%%%%%%%%%%
\subsection{Particle Yields}
\label{ssec_chem}
%%%%%%%%%%%%%%%%%%%%%%%%%%%%%%%%%%%%%%%%%%%%%%%%%%%%%
The PCE part of our EoS ensures that the stable-hadron numbers are
approximately conserved between chemical and thermal freezeout. In
Tab.~\ref{yieldtable} we compare our $\pi$, $K$ and $p$
multiplicities in central Au-Au collisions, calculated at $T_{\rm
fo}$, to STAR~\cite{Abelev:2008ez} and PHENIX~\cite{Adler:2003cb}
data. Since we employ the chemical freezeout temperature we adjusted
the total entropy to obtain the central value of the proton number
measured by STAR (including weak feeddown); consequently, the
calculated $\pi$ (corrected for weak feeddown), $K$ and $\bar p$
yields are within the experimental errors (residual deviations
inside the errors may be caused, e.g., by slight variations in the
hadronic resonances included in the EoS). Good agreement is also
found with the kaon measurements of PHENIX, while the calculated
pion yields are slightly above the 1-$\sigma$ upper limit of the
PHENIX pions (which include weak feeddown). A more significant
discrepancy arises with the PHENIX anti-/protons (weak feeddown
corrected). Similar findings are reported in other hydro-model
fits~\cite{Huovinen:2007xh,pasi11}. We note that a slightly higher
centrality selection in the PHENIX data could still be compatible
with our calculated  kaons while improving the agreement with pions
and anti-/protons.
%After subtracting hyperon ($\Lambda, \Sigma,
%\Xi$ and $\Omega$) weak feeddown, the proton and anti-proton yields
\begin{table*}
% For LaTeX tables use
\begin{tabular}{lcccccc}
\hline\noalign{\smallskip}
 $\mathrm{d}N/\mathrm{d}y$
                      & $\pi^+$ & $\pi^-$ & $K^+$ & $K^-$ & $p$ & $\bar{p}$\\
\noalign{\smallskip}\hline\noalign{\smallskip} STAR  & $322\pm 25$
                                & $327\pm 25$
                                          & $51.3\pm 6.5$
                                                  & $49.5\pm 6.2$
                                                          & $34.7\pm 4.4$
                                                             & $26.7\pm 3.4$ \\

PHENIX
                      & $286.4\pm 24.2$
                                & $281.8\pm 22.8$
                                          & $48.9\pm 5.2$
                                                  & $45.7\pm 5.2$
                                                          & $18.4\pm 2.6$
                                                             & $13.5\pm 1.8$ \\
$Y$ feeddown included     & 312.2 & 314.4  & 48.2   & 48.4    &  34.0 &  26.2      \\
$Y$ feeddown subtracted   & 303.1 & 303.1  &  48.1  &  48.2   & 25.8
&19.9\\
\noalign{\smallskip}\hline
\end{tabular}
\caption{Comparison of the calculated bulk-particle yields with STAR
and PHENIX measurements at midrapidity in 0-5\% central Au-Au
collisions at $\sqrt{s_{\mathrm{NN}}}=200$ GeV. Results are shown
with and without feeddown from hyperon decays. The initial entropy
has been adjusted to reproduce the observed STAR
anti-/protons~\cite{Abelev:2008ez}. The STAR pions are weak feeddown
corrected, while the protons and anti-protons include hyperon
feeddown. PHENIX anti-/protons are weak feeddown
corrected~\cite{Adler:2003cb}, but not the pions.}
\label{yieldtable}
\end{table*}

%%%%%%%%%%%%%%%%%%%%%%%%%%%%%%%%%%%%%%%%%%%%%%%%%%%%%
\subsection{Single-Particle Spectra}
\label{ssec_pt}
%%%%%%%%%%%%%%%%%%%%%%%%%%%%%%%%%%%%%%%%%%%%%%%%%%%%%
\begin{figure}[!t]
\includegraphics[width=\columnwidth]{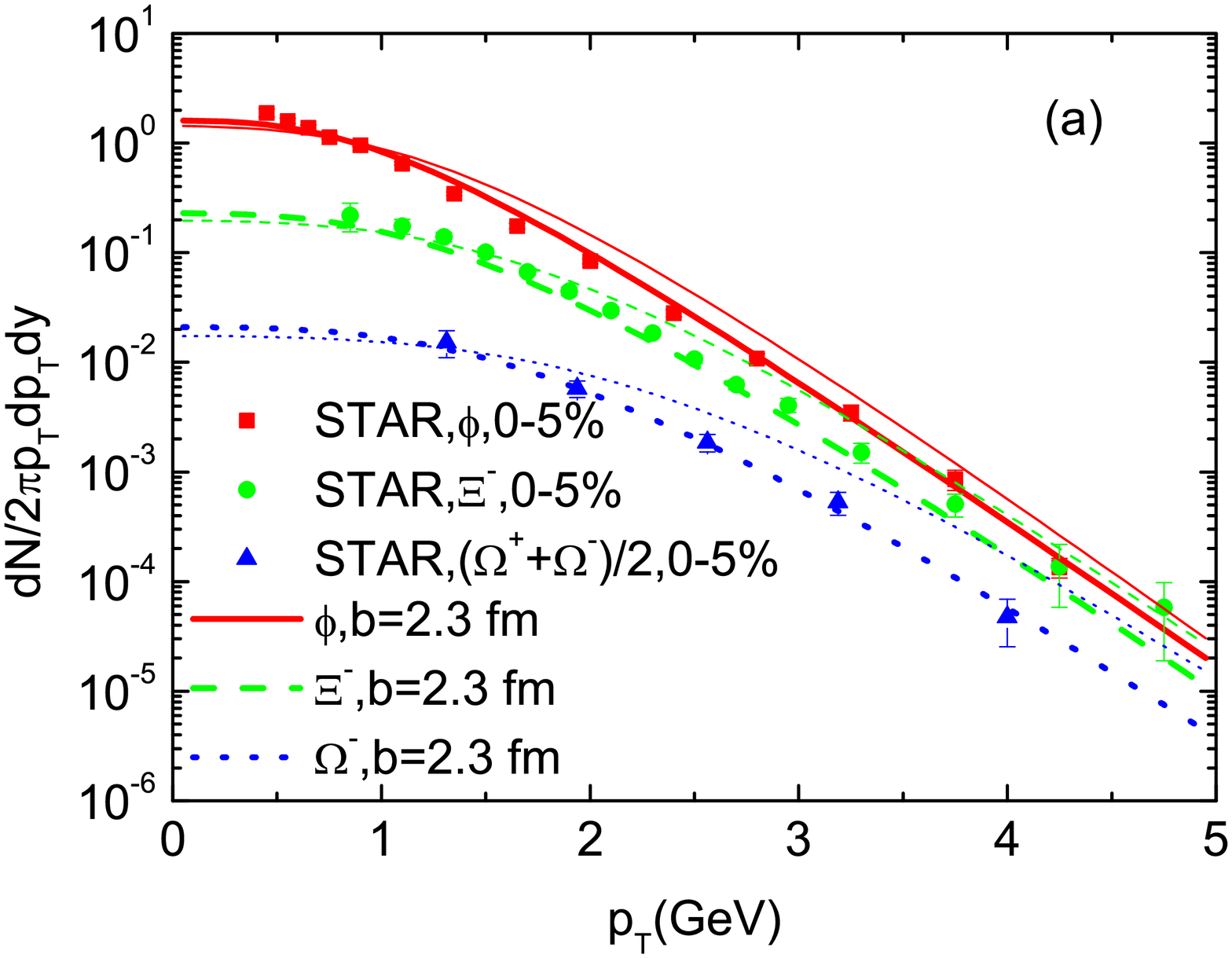}
\includegraphics[width=\columnwidth]{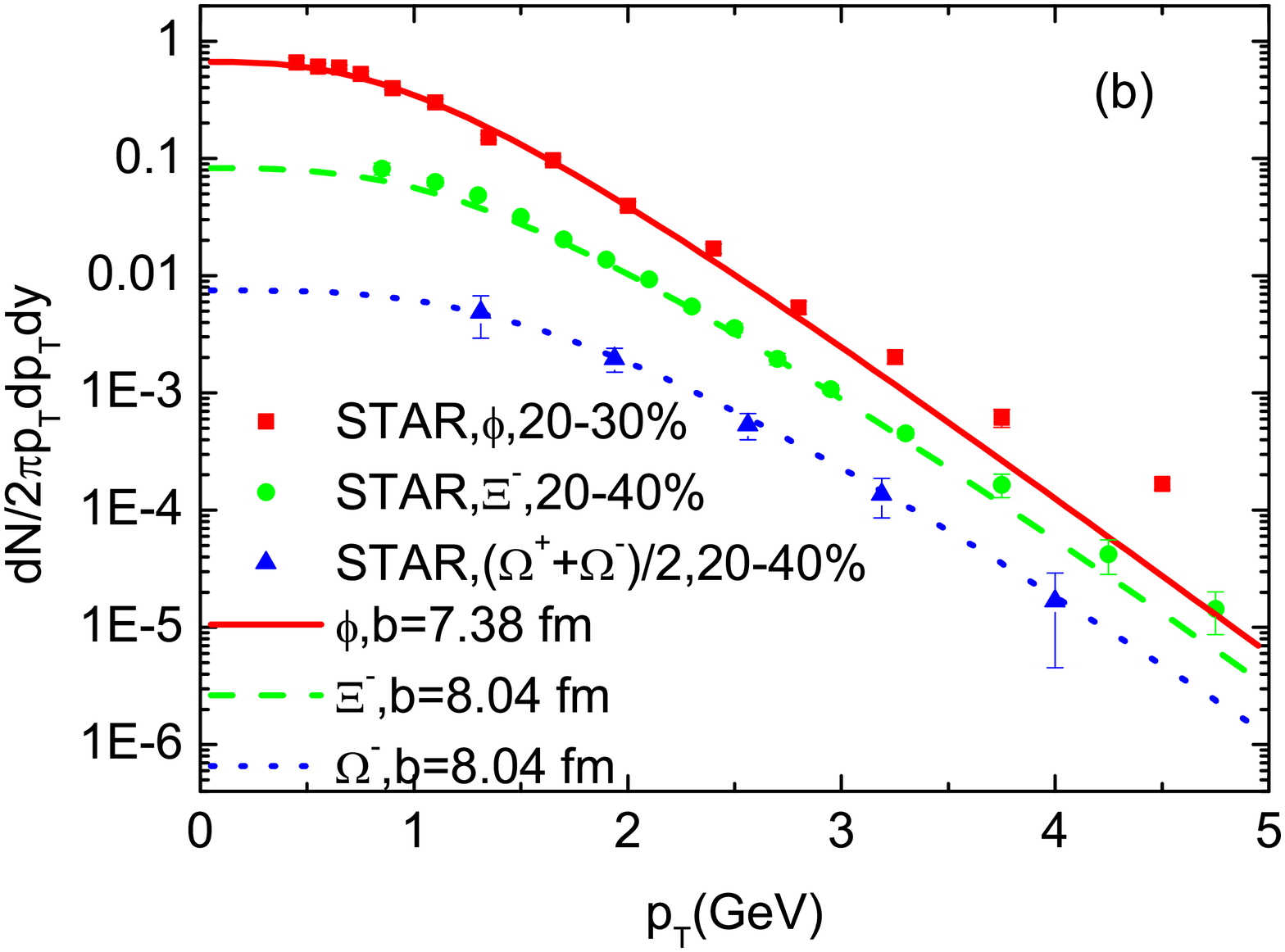}
\caption{(Color online) (a) Our calculated $p_T$-spectra of
multistrange particles ($\phi$, $\Xi$ and $\Omega$) in 0-5\% central
Au-Au collisions compared to STAR
data~\cite{Adams:2006ke,Abelev:2007rw}; the thick (thin) lines
correspond to kinetic freezeout at $T_{\rm ch}=160$\,MeV ($T_{\rm
fo}=110$\,MeV, with identical normalization). (b) The same
comparison for semicentral collisions (without thin lines for late
freezeout).} \label{multisrage_spectra}
\end{figure}
\begin{figure}[!t]
\includegraphics[width=\columnwidth]{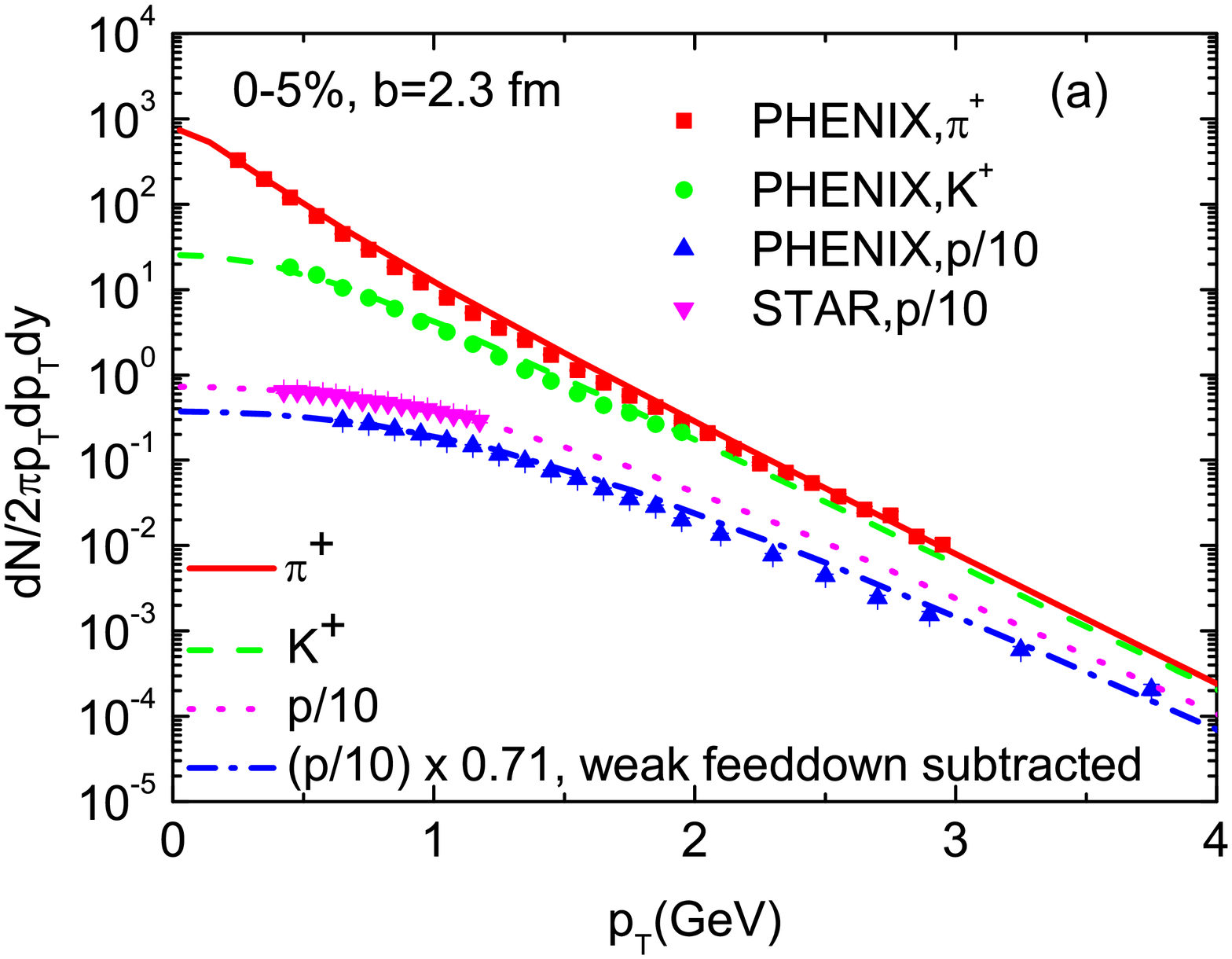}
\includegraphics[width=\columnwidth]{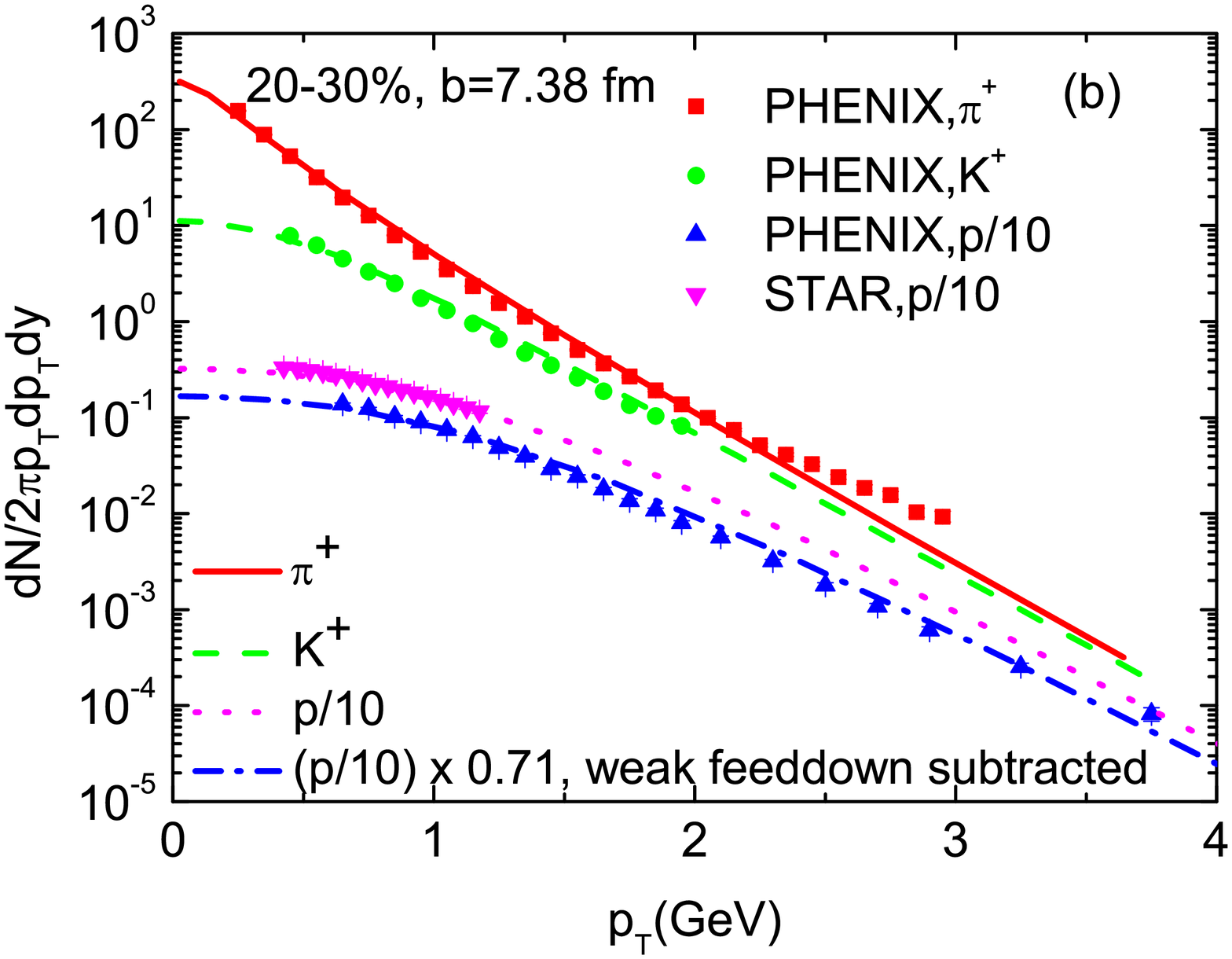}
\caption{(Color online) (a) Our calculated $p_T$ spectra of bulk
particles ($\pi^+$, $K^+$ and $p$) in 0-5\% central Au-Au collisions
compared to PHENIX~\cite{Adler:2003cb} and STAR
data~\cite{Adams:2003xp}. (b) The same comparison for semicentral
collisions.} \label{bulk_spectra}
\end{figure}

Let us first turn to the multistrange hadron spectra ($\phi$, $\Xi$
and $\Omega$) which we evaluate at the chemical-freezeout
temperature, $T_{\rm ch}=160~{\rm MeV}$, where the energy density is
$e_{\rm ch}=0.372~{\rm GeV/fm^3}$. As an additional centrality class
we consider $20-40\%$ central Au-Au collision which we approximate
with $b=8.04~{\rm fm}$ with an initial-entropy density in the center
of the transverse plane of $s_0=130.9~{\rm fm^{-3}}$ to reproduce
the pertinent bulk-particle yields; all other parameters are the
same as for $20-30\%$ centrality (including the strangeness
suppression factor of $\gamma_s=0.85$~\cite{Abelev:2008ez}). The
comparison of our calculated multistrange hadron spectra to STAR
measurements are shown in Fig.~\ref{multisrage_spectra}, showing
good agreement in both central and semicentral collisions. As
typical for hydrodynamic simulations, the description extends to
slightly higher $p_T$ for higher centrality, here up to
$p_T\simeq4-5$\,GeV for hyperons.

With the same set of parameters, the freezeout of bulk particles,
($\pi$, $K$, $p$) is evaluated at  $T_{\rm fo}=110~{\rm MeV}$ (at an
energy density of $e_{\rm fo}=0.1094~{\rm GeV/fm^3}$). This is
consistent with values extracted from systematic blast-wave fits
performed by the experimental collaborations to their
data~\cite{BurwardHoy:2002xu,Abelev:2008ez}. Consequently, our
calculated spectra agree well with their spectra as well,
cf.~Fig.~\ref{bulk_spectra} albeit with absolute normalization owing
to the implementation of chemical freezeout.
%As found for
%$\phi$, the more central the collision, the larger $p_T$ range in
%which the measured pion spectrum can be fairly well reproduced.
One exception are the PHENIX protons (including weak feeddown
corrections from hyperons): while the spectral shape is well
described, a renormalization factor of 0.71 needs to be applied for
an optimal description of the absolute yields (in accord with
discussion in Sec.~\ref{ssec_chem} and Tab.~\ref{yieldtable}).

In Ref.~\cite{Huovinen:2007xh}, which employs an EoS similar to that
used in AZHYDRO, it was deduced that bulk-particle spectra can be
fitted better with a chemical freezeout temperature of 150\,MeV,
ca.~10\,MeV lower than the value suggested by statistical model
fits~\cite{BraunMunzinger:2003zd,Becattini:2000jw,Abelev:2008ez}.
Our results indicate that with a modern lattice EoS and initial flow
the chemical freezeout temperature in hydrodynamics can be made
compatible with statistical model fits.

To illustrate the significance of the early freezeout of
multistrange hadrons in our calculations, we also plot their spectra
at $T_{\rm fo}=110$\,MeV in central Au-Au (thin lines in the upper
panel of Fig.~\ref{multisrage_spectra}). One sees that the
additional radial flow developed in the hadronic phase hardens the
spectra substantially leading to a systematic overprediction of the
data with increasing $p_T$. The discrepancy is largest for the
$\Omega^-$ and less pronounced for the $\Xi$ and $\phi$. The latter
may not be surprising; e.g., the $\Xi$ possesses one light valence
quark and at least one pion-resonance excitation ($\Xi(1530)$),
while the $\phi$ couples strongly to both $\pi\rho$ and $K\bar K$
channels. Thus both $\phi$ and $\Xi$ might develop a reaction rate
in hadronic matter which allows them to pick up some additional
collectivity after chemical freezeout, leading to an effective
freezeout temperature slightly below $T_{\rm ch}$; similar findings
are reported in Ref.~\cite{Hirano:2005xf} for the $\phi$.

%%%%%%%%%%%%%%%%%%%%%%%%%%%%%%%%%%%%%%%%%%%%%%%%%%%%%
\subsection{Elliptic Flow}
\label{ssec_v2}
%%%%%%%%%%%%%%%%%%%%%%%%%%%%%%%%%%%%%%%%%%%%%%%%%%%%%
\begin{figure}[!t]
\includegraphics[width=\columnwidth]{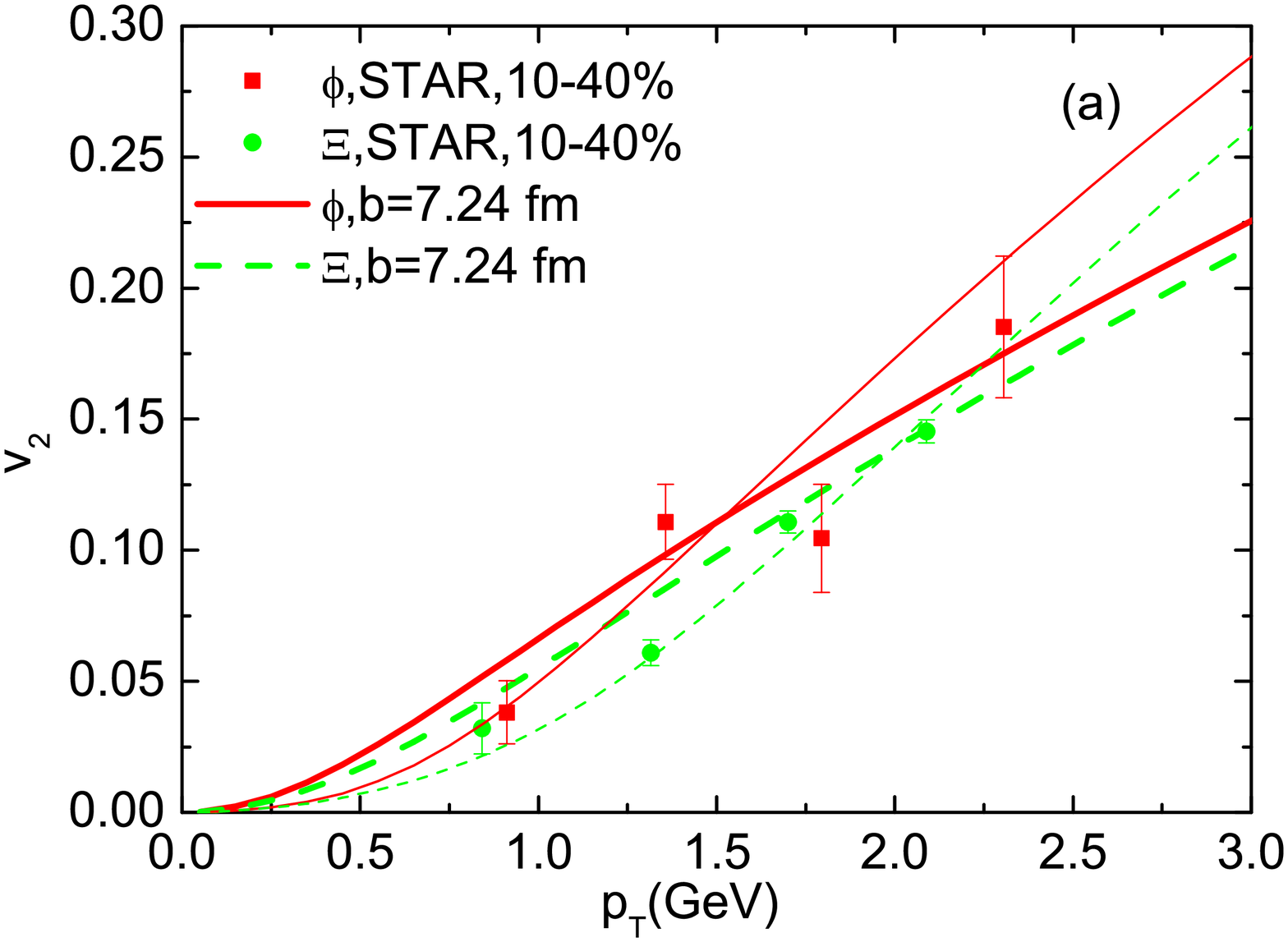}
\includegraphics[width=\columnwidth]{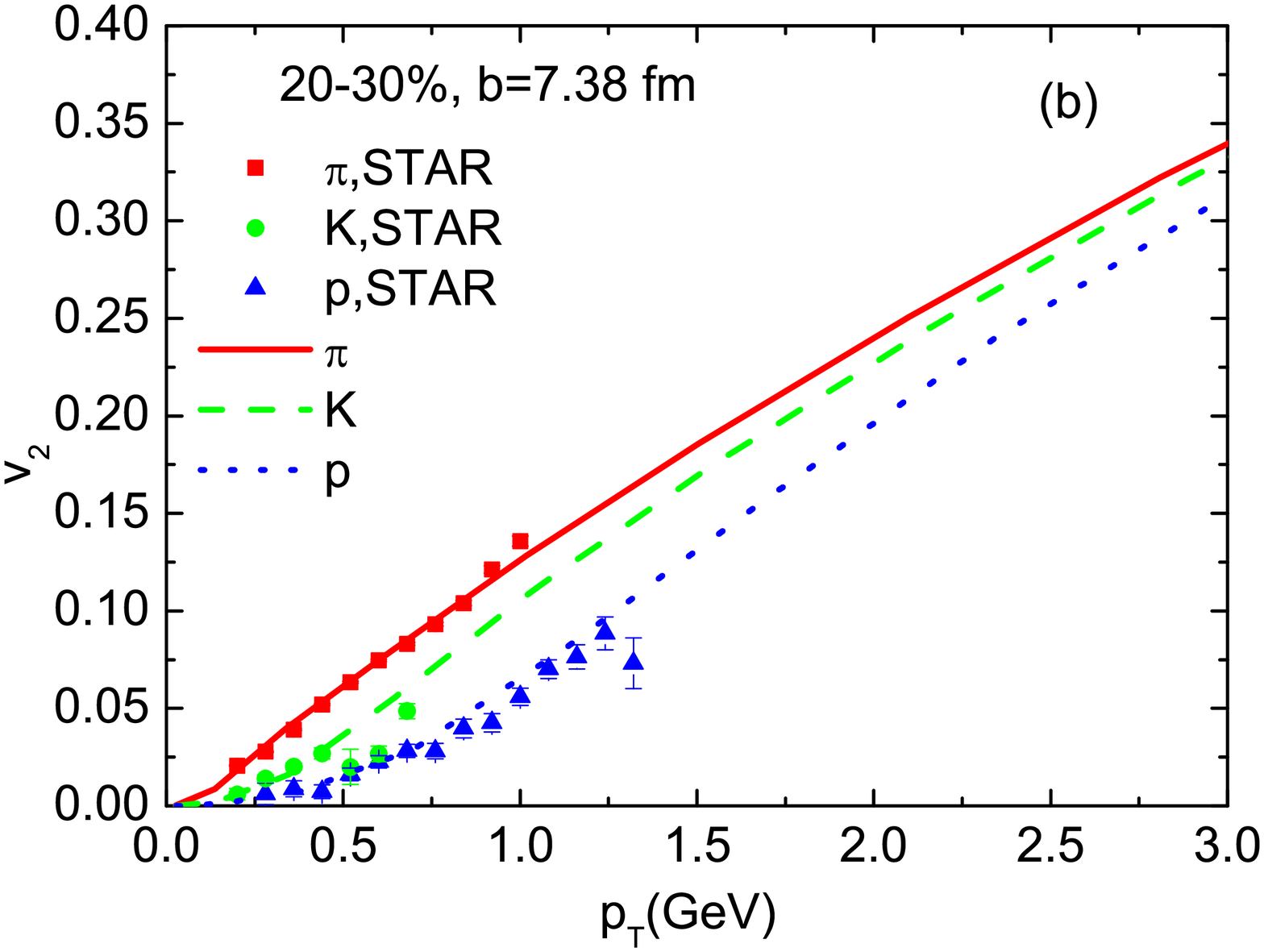}
\caption{(Color online) (a) Our calculations for the elliptic-flow
coefficient as a function of $p_T$ for $\phi$ and $\Xi$ compared to
STAR data~\cite{Abelev:2007rw,Abelev:2008ed} in 10-40\% central
Au-Au collisions; the thick (thin) lines correspond to kinetic
freezeout at $T_{\rm ch}=160$\,MeV ($T_{\rm fo}=110$\,MeV, with
identical normalization); the calculations employ the same impact
parameter as for the 20-30\% centrality class. (b) The same
comparison for $\pi$, $K$ and $p$ with STAR data for 20-30\%
centrality from Ref.~\cite{Adams:2004bi}; only the results for
$T_{\rm fo}=110$\,MeV are shown.} \label{v2}
\end{figure}
We finally turn to the elliptic flow, which in our calculation is
controlled, to a limited extent, by the parameters $\alpha_2$ and
$\kappa$ characterizing the anisotropy of initial flow, and, to a
lesser extent, by the interplay with the initial radial flow (the
initial spatial anisotropy is fixed by our assumption of the $n_{\rm
BC}$ profile of the entropy density). As is well
known~\cite{Kolb:2003dz,Teaney:2001av,Huovinen:2006jp,Huovinen:2001cy},
the $v_2$ of massive particles is reduced at low $p_T$ by a large
radial flow. This, in particular, also occurs when initial flow
fields are introduced~\cite{Kolb:2002ve}, thus mitigating the
problem of a growing pion $v_2$ in the hadronic phase when a PCE EoS
is employed~\cite{Kolb:2002ve,Hirano:2005wx,Huovinen:2007xh}.

Our results for the anisotropy coefficient of multistrange and bulk
particles in semicentral Au-Au collisions are compared to STAR data
in the upper and lower panel of  Fig.~\ref{v2}, respectively. Our
fit yields fairly good agreement for both hadron classes. Clearly,
the aforementioned problem of previous PCE implementations is
largely resolved due to our more explosive expansion, in connection
with initial flow fields, which suppress a significant increase of
$v_2$ in the late stage of the hydrodynamic evolution. This was
already indicated by the saturation of the anisotropy of the
energy-momentum tensor anisotropy in Fig.~\ref{bulk_evolution}.

To illustrate the significance of sequential freezeout, we display
again the $v_2$ of $\phi$ and $\Xi$ at $T_{\rm fo}=110$\,MeV by thin
lines in the upper panel of Fig.~\ref{v2}. Compared to the $p_T$
spectra (the upper panel of Fig.~\ref{multisrage_spectra}), the
current $v_2$ data are less discriminatory for the freezeout
temperature of multistrange hadrons. In accordance with the remarks
at the end of Sec.~\ref{ssec_pt}, the $v_2$ of $\Xi$ may also favor
a kinetic-freezeout temperature slightly below $T_{\rm ch}$.

%%%%%%%%%%%%%%%%%%%%%%%%%%%%%%%%%%%%%%%%%%%%%%%%%%%%%%%%%%%%%%
\section{Summary and Conclusion}
\label{summary}
%%%%%%%%%%%%%%%%%%%%%%%%%%%%%%%%%%%%%%%%%%%%%%%%%%%%%%%%%%%%%%
In the present study we have explored the capability of ideal
hydrodynamics to simultaneously and quantitatively describe bulk-
and multistrange-hadron spectra and $v_2$ in Au-Au collisions at
RHIC. Specifically, we have augmented an existing 2+1D ideal hydro
code by (i) an equation of state compatible with recent lattice QCD
data matched to a hadron resonance gas in partial chemical
equilibrium, (ii) a sequential kinetic freezeout of multistrange and
bulk particles, and (iii) a compact initial density profile with
non-zero radial and elliptic flow. We deem these amendments
``reasonable" in the sense of being either suggested by theory
(lattice EoS), experiment (chemical freezeout) and empirical fits
(early kinetic freezeout of multistrange hadrons), or at least
plausible (initial flow and compact profiles). The above items also
encompass three of the four components identified for solving the
pion HBT problem. The main practical consequences of the
modifications are a more rapid build-up of the radial and elliptic
flow, where the latter essentially levels off after hadronization.
Phenomenologically, the underlying parameters (basically the three
in the initial-flow field parametrization) can be adjusted as to
render multistrange hadrons' kinetic freezeout at $T_{\rm ch}$
compatible with data, and to subsequently avoid an overshooting of
the bulk-particle $v_2$ at $T_{\rm fo}$. We consider this a
significant improvement over existing ideal-hydro calculations,
thereby corroborating the empirical picture of early freezeout of
multistrange hadrons.

We did not include the effects of viscosities nor of initial-state
fluctuations, which are both clearly needed for realistic
hydrodynamic simulations of heavy-ion collisions. However, we
believe that our study can still serve as a useful baseline for the
effects we have included. In addition, we anticipate that our amended
AZHYDRO can be a valuable tool as a realistic fireball background
for evaluating heavy-quark(onium) and electromagnetic probes.

\vspace{0.3cm}

\acknowledgments MH acknowledges useful correspondence with C.~Shen
and E.~Frodermann. We thank U.~Heinz and C.~Shen for valuable comments
on the manuscript. This work was supported by the U.S. National
Science Foundation (NSF) through CAREER grant PHY-0847538 and grant
PHY-0969394, by the A.-v.-Humboldt Foundation, by the RIKEN/BNL
Research Center and DOE grant DE-AC02-98CH10886, and by the JET
Collaboration and DOE grant DE-FG02-10ER41682.

\end{document}